\def\ra{\rangle}

\def\be{\begin{equation}}
\def\ee{\end{equation}}
\def\ba{\begin{array}}
\def\ea{\end{array}}

\documentclass[prl,showpacs,twocolumn,amsmath]{revtex4}
\usepackage{epsfig,graphicx}
\usepackage{amsmath}
\input amssym.def

\begin{document}
\title{Measure and Detection of Genuine Multipartite Entanglement for Tripartite Systems}
\author{Ming Li$^{1}$}
\author{Lingxia Jia$^{1}$}
\author{Jing Wang$^{1}$}
\author{Shuqian Shen$^{1}$}
\author{Shao-Ming Fei$^{2,3}$}
\affiliation{$^1$College of the Science, China University of
Petroleum, 266580 Qingdao, China\\
$^2$School of Mathematical Sciences, Capital Normal University, Beijing 100048, China\\
$^3$ Max-Planck-Institute for Mathematics in the Sciences, 04103
Leipzig, Germany}

\begin{abstract}
It is a computationally hard task to certify genuine multipartite entanglement (GME). We investigate the relation between the
norms of the correlation vectors and the detection of GME for tripartite quantum systems. A sufficient condition for GME and an effective lower bound for the GME concurrence are derived. Several examples are considered to show the effectiveness of the criterion and the lower bound of GME concurrence.

\end{abstract}

\smallskip

 \maketitle

\section{Introduction}

Quantum entanglement is a remarkable resource in the
theory of quantum information, with various
applications \cite{nielsen, di}.
A multipartite quantum state
that is not separable with respect to any bi-partition is
said to be genuine multipartite entangled \cite{guhnerev}.
As one of the
important type of entanglement, genuine multipartite entanglement(GME) offers significant
advantage in quantum tasks comparing with bipartite entanglement
\cite{mule1}. In particular, it is the basic ingredient in
measurement-based quantum computation \cite{mule2}, and is
beneficial in various quantum communication protocols \cite{mule3},
including secret sharing \cite{mule4} (cf. \cite{mule5}), extreme spin squeezing \cite{srensen}, high sensitivity in some
general metrology tasks \cite{hyllus}, quantum computing using
cluster states \cite{rauss}, and multiparty quantum network \cite{hillery,scarani,zhao}.
However, detecting and measuring quantum entanglement turn out to be quite
difficult.
To detect GME, series of linear and nonlinear entanglement witnesses \cite{12, huber,
vicente3, huber1, wu, sperling, 14, 15, claude, horo}, generalized
concurrence for GME \cite{ma1,
ma2,gaot1,gaot2}, and Bell-like inequalities \cite{bellgme} were derived and a characterization in terms of semi-definite programs
(SDP) was developed \cite{10}.
Nevertheless, the problem remains far from being satisfactorily solved.

From the norms of the correlation tensors in the generalized Bloch
representation of a quantum state, separable conditions for both bi- and multi-partite quantum states
have been presented in \cite{vicente1,vicente2,hassan,ming}. A multipartite
entanglement measure for N-qubit and N-qudit pure states is given in
\cite{hassan1,hassan2}. A general framework for detecting genuine multipartite entanglement
and non full separability in multipartite quantum systems of
arbitrary dimensions has been introduced in \cite{vicente3}.
In \cite{mingbell} it has been shown that the norms of the correlation tensors has
a close relationship to the maximal violation of a kind of multi Bell inequalities and to the concurrence \cite{GE}.

In this paper, we investigate the genuine tripartite entanglement in terms
of the norms of the correlation tensors and GME concurrence for tripartite qudit quantum systems. We derive criteria to detect GME. An
effective lower bound for GME concurrence is also presented.

\section{Criteria for detecting GME}

In this section, we present a criterion to detect GME by using the approach presented in \cite{vicente3} for tripartite qudit systems. We start with some definitions and notations.

Let $H_i^d$, $i=1,2,3$, denote $d$-dimensional Hilbert spaces. A tripartite state $\rho  \in H_1^d \otimes H_2^d \otimes H_3^d$ can be expressed as $\rho  = \sum {{p_\alpha }} \left| {{\psi _\alpha }} \right\rangle \left\langle {{\psi_\alpha }} \right|$, where $0<p_\alpha\leq 1$,
$\sum {{p_\alpha }}  = 1$, $\left| {{\psi _\alpha }} \right\rangle \in H_1^d \otimes H_2^d \otimes H_3^d$ are normalized pure states.
If all $\left| {{\psi _\alpha }} \right\rangle$ are biseparable, namely, either
$\left| {{\psi _\alpha }} \right\rangle = \left| {\varphi _\alpha ^1} \right\rangle  \otimes \left| {\varphi _\alpha ^{23}} \right\rangle $ or $\left| {{\psi _\beta }} \right\rangle  = \left| {\varphi _\beta ^2} \right\rangle  \otimes \left|{\varphi _\beta ^{13}} \right\rangle $ or $\left| {{\psi _\gamma }} \right\rangle  = \left| {\varphi _\gamma ^3} \right\rangle  \otimes \left|{\varphi _\gamma ^{12}} \right\rangle$,
where $\left| {\varphi_\gamma^i} \right\rangle$ and $\left| {\varphi_\gamma^{ij}} \right\rangle$ denote pure states in $H_i^d$ and $H_i^d \otimes H_j^d$ respectively,
then $\rho $ is said to be bipartite separable. Otherwise, $\rho $ is called genuine multipartite entangled.

Let $\lambda_i$, $i =1,\cdot  \cdot  \cdot, {d^2} - 1$, denote the generators of the special unitary group $SU(d)$\cite{lambda}, and $I$ the $d\times d$ identity matrix .
Any $\rho  \in H_1^d \otimes H_2^d \otimes H_3^d$ can be represented as follows:
\begin{widetext}
\begin{eqnarray}\label{rho}
\rho  &=& \frac{1}{{{d^3}}}I \otimes I \otimes I + \frac{1}{{2d^2}}(\sum {t_i^1{\lambda _i} \otimes I \otimes I}  + \sum {t_j^2I \otimes {\lambda _j} \otimes I}  + \sum {t_k^3I \otimes I \otimes {\lambda _k}} )\nonumber\\
 &&+ \frac{1}{{4d}}(\sum {t_{ij}^{12}{\lambda _i} \otimes {\lambda _j} \otimes I} {\kern 1pt} {\kern 1pt} {\kern 1pt}  + \sum {t_{ik}^{13}{\lambda _i} \otimes I \otimes {\lambda _k}}  + \sum {t_{jk}^{23}I \otimes {\lambda _j} \otimes {\lambda _k}} )+  \frac{1}{8}\sum {t_{ijk}^{123}{\lambda _i} \otimes {\lambda _j} \otimes {\lambda _k},}
\end{eqnarray}
\end{widetext}
where ${t_{i}^{1}} = tr\left( {\rho\, {\lambda _i} \otimes I \otimes I} \right)$, ${t_{j}^{2}} = tr\left( {\rho\, I\otimes {\lambda _j} \otimes I} \right)$,
${t_{k}^{3}} = tr\left( {\rho\, I\otimes I \otimes {\lambda _k}} \right)$, ${t_{ij}^{12}} = tr\left( {\rho \,{\lambda _i} \otimes {\lambda _j} \otimes I} \right)$.
${t_{ik}^{13}} = tr\left( {\rho {\lambda _i} \otimes I \otimes {\lambda _k}} \right)$, ${t_{jk}^{23}} = tr\left( {\rho\, I \otimes {\lambda _j} \otimes {\lambda _k}} \right)$ and
${t_{ijk}^{123}} = tr\left( {\rho \,{\lambda _i} \otimes {\lambda _j} \otimes {\lambda _k}} \right)$.
In the following, we set
$T^{(1)}$, $T^{(2)}$, $T^{(3)}$, $T^{(12)}$, $T^{(13)}$, $T^{(23)}$ and $T^{(123)}$ to be the vectors (tensors) with entries $t_{i}^{1}$, $t_{j}^{2}$, $t_{k}^{3}$, $t_{ij}^{12}$, $t_{ik}^{13}$, $t_{jk}^{23}$ and $t_{ijk}^{123}$, $i, j, k=1, 2, \cdots, d^2-1$, which are the so called correlation vectors.

Let ${\left\| M \right\|_k} = \sum\limits_{i = 1}^k {{\sigma _i}}$ denote the $k$-norm for an $n \times n$ matrix $M$, where ${\sigma _i}$, $i=1,...,n$, are the singular values of $M$ in decreasing order.
${\left\| M \right\|_n} = {\left\| M \right\|_{KF}}$ is just the Key-Fan norm. Denote $\left\| \cdot \right\|$ the Frobenius norm of a vector or a matrix.
Let ${T_{\underline 1 23}}$, ${T_{\underline 2 1 3}}$ and ${T_{\underline 3 12 }}$ be the matrices with entries $t_{i,d(j - 1) + k} = {t_{ijk}}$, $t_{j,d(i - 1) + k} = {t_{ijk}}$ and $t_{k,d(i - 1) + j} = {t_{ijk}},$ respectively. Set ${M_k}(\rho ) = \frac{1}{3}({\left\| {{T_{\underline 1 23}}} \right\|_k} + {\left\| {{T_{\underline 2 1 3}}} \right\|_k} + {\left\| {{T_{\underline 3 12 }}} \right\|_k}).$

\textbf{\textit{Lemma:}} \label{lema}
For a pure tripartite qudit state, we have for any $k=1,2, \cdots, d^2-1$ and $\underline j l m=\underline 1 2 3, \underline 2 1 3, \underline 3 1 2$ that \\
(i) if the state is fully separable, then
\be{\left\| {{T_{\underline j lm}}} \right\|_k} \le \sqrt {\frac{{8{{(d - 1)}^3}}}{{{d^3}}}} ;\ee
(ii) if the state is separable under bipartite partition $j\left| {lm} \right.$, then
\be{\left\| {{T_{\underline j lm}}} \right\|_k} \le \sqrt {\frac{{8{{(d - 1)}^2}(d + 1)}}{{{d^3}}}} ;\ee
(iii) if the state is entangled under bipartite partition $j\left| {lm} \right.$, then
\be{\left\| {{T_{\underline j lm}}} \right\|_k} \le \sqrt {\frac{{8k{{(d - 1)}^2}(d + 1)}}{{{d^3}}}} .\ee

\textbf{Proof.} We shall use repeatedly
$\left\| {{T^{(i)}}} \right\| \le \sqrt {\frac{{2(d - 1)}}{d}}, i=1,2,3, \left\| {{T^{(lm)}}} \right\| \le 2\sqrt {\frac{{{d^2} - 1}}{{{d^2}}}}, lm=12,13,23$ \cite{vicente3}
and ${\left\| M \right\|_k} \le \sqrt k \left\| M \right\|$ for any matrix $M$. Then we have\\
(i)
\begin{align*}
{\left\| {{T_{\underline j lm}}} \right\|_k} &= {\left\| {({T^{(j)}}) \cdot {{({T^{(l)}} \otimes {T^{(m)}})}^t}} \right\|_k}\\
& = \left\| {{T^{(j)}}} \right\|\left\| {{T^{(l)}} \otimes {T^{(m)}}} \right\| = \left\| {{T^{(j)}}} \right\|\left\| {{T^{(l)}}} \right\|\left\| {{T^{(m)}}} \right\|\\
 &\le \sqrt {\frac{{8{{(d - 1)}^3}}}{{{d^3}}}};
\end{align*}
(ii)
\begin{align*}
{\left\| {{T_{\underline j lm}}} \right\|_k} & = {\left\| {({T^{(j)}}) \cdot ({T^{(lm)}})}^t \right\|_k} \\
&   = \left\| {{T^{(j)}}} \right\|\left\| {{T^{(lm)}}} \right\| \le \sqrt {\frac{{8{{(d - 1)}^2}(d + 1)}}{{{d^3}}}};
\end{align*}
(iii)
\begin{align*}
{\left\| {{T_{\underline j lm}}} \right\|_k} &= {\left\| {({T^{(\underline j l)}}) \otimes {{({T^{(m)}})}^t}} \right\|_k}\\
&   = {\left\| {{T^{(\underline j l)}}} \right\|_k}\left\| {{T^{(m)}}} \right\| \le \sqrt {\frac{{8k{{(d - 1)}^2}(d + 1)}}{{{d^3}}}};
\end{align*}
where we have denoted $T^{(\underline j l)}$ the matrix with entries ${t_{xy}^{jl}}$.

\textit{\textbf{Theorem 1:} If for a tripartite qudit state $\rho $, it holds that
\be\label{T1}
M_k(\rho ) > \frac{2\sqrt{2}}{3}(2\sqrt{k}+1)\frac{d-1}{d}\sqrt{\frac{d+1}{d}}
\ee
for any $k=1,2,\cdots, d^2-1$, then $\rho $ is genuine multipartite entangled.}

{\textbf{Proof.}} Assume that $\rho$ is bipartite separable. By using the above Lemma, we get
\begin{widetext}
\begin{eqnarray*}
{M_k}(\rho )&=&\frac{1}{3}({\left\| T_{\underline 1 23}(\rho) \right\|_k} + {\left\| T_{\underline 2 13}(\rho) \right\|_k} + {\left\| T_{\underline 3 12} (\rho) \right\|_k})\\
&=&\frac{1}{3}({\left\|\sum_{\alpha}p_{\alpha} T_{\underline 1 23}(|\psi_{\alpha}\rangle) \right\|_k} + {\left\|\sum_{\alpha}p_{\alpha} T_{\underline 2 13}(|\psi_{\alpha}\rangle) \right\|_k} + {\left\|\sum_{\alpha}p_{\alpha} T_{\underline 3 12}(|\psi_{\alpha}\rangle) \right\|_k})\\
&\leq&\frac{1}{3}\sum_{\alpha}p_{\alpha} ({\left\|T_{\underline 1 23}(|\psi_{\alpha}\rangle) \right\|_k} + {\left\| T_{\underline 2 13}(|\psi_{\alpha}\rangle) \right\|_k} + {\left\|T_{\underline 3 12}(|\psi_{\alpha}\rangle) \right\|_k})\\
&\leq&\frac{1}{3}\sum_{\alpha}p_{\alpha}(2\sqrt {\frac{{8k{{(d - 1)}^2}(d + 1)}}{{{d^3}}}}  + \sqrt {\frac{{8{{(d - 1)}^2}(d + 1)}}{{{d^3}}}} ) \\
&=&\frac{1}{3}(2\sqrt {\frac{{8k{{(d - 1)}^2}(d + 1)}}{{{d^3}}}}  + \sqrt {\frac{{8{{(d - 1)}^2}(d + 1)}}{{{d^3}}}} )\\
&=&\frac{2\sqrt{2}}{3}(2\sqrt{k}+1)\frac{d-1}{d}\sqrt{\frac{d+1}{d}}.
\end{eqnarray*}
\end{widetext}
Thus once the inequality is violated, the quantum state $\rho$ must be genuine multipartite entangled.
\hfill \rule{1ex}{1ex}

{\it{Remark.}} For $d = 2$, $k = 3$, Theorem 1 reduces to the Theorem 2 in \cite{vicente3}. If we set $d = 2$, and $k = 1,2$ one finds that Theorem 1 is strictly covered by Theorem 2 in \cite{vicente3}. However, our theorem can detect GME for any tripartite qudit systems rather than only tripartite qubit systems.

\textit{\textbf{Example 1:}}
Consider quantum state $\rho  \in H_1^3 \otimes H_2^3 \otimes H_3^3,$ $\rho  = \frac{1-x}{27}I + x\left| \varphi  \right\rangle \left\langle \varphi  \right|,$ where $\left| \varphi  \right\rangle  = \frac{1}{{\sqrt 3 }}(\left| {000} \right\rangle  + \left| {111} \right\rangle  + \left| {222} \right\rangle )$ is the GHZ state. By Theorem 1 in \cite{vicente3} we can detect GME for $0.894427<x\leq 1$. With our Theorem 1, by setting $d=3$ and $k=8$(which is the optimal selection) one detects GME for $0.716235<x\leq 1$(see Fig. 1).
\begin{figure}[h]
\begin{center}
\resizebox{6cm}{!}{\includegraphics{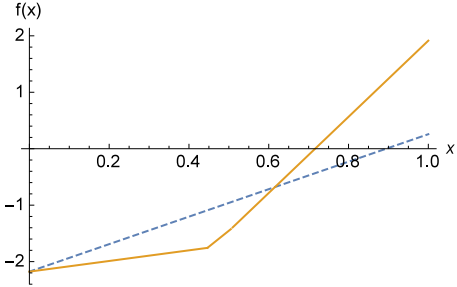}}
\end{center}
\caption{Vicente criterion (dashed line) v.s. the lower bound (\ref{T1}) in this manuscript (solid line). We have used $f(x)$ to denote the difference between the left and the right side of the two criteria. From the figure one sees that $\rho$ is genuine multipartite entangled for $0.894427<x\leq 1$ by Vicente's result with $f(x)=-2.177324 + 2.434322x$. One computes $f(x)=-4.83138 + 6.74552x$ by our proposition. Thus GME is detected for $0.716235<x\leq 1$.\label{fig1}}
\end{figure}

\section{Lower bound of GME concurrence}

The GME concurrence is proved a well defined measure \cite{ma1,ma2}.
For a pure state $|\psi\ra\in H_1^d \otimes H_2^d \otimes H_3^d$, the GME concurrence is defined by
\begin{eqnarray*}
C_{GME}(|\psi\ra)=\sqrt{\min\{1-tr(\rho_1^2),1-tr(\rho_2^2),1-tr(\rho_3^2)\}},
\end{eqnarray*}
where $\rho_i$ is the reduced matrix for the $i$th subsystem.
For mixed state $\rho\in H_1^d \otimes H_2^d \otimes H_3^d$, the GME concurrence is then defined by the convex roof
\begin{eqnarray}
C_{GME}(\rho)=\min\sum_{\{p_{\alpha},|\psi_{\alpha}\ra\}}p_{\alpha}C_{GME}(|\psi_{\alpha}\ra).
\end{eqnarray}
The minimum is taken over all pure ensemble decompositions of $\rho$.

Since one has to find the optimal ensemble to carry out the minimization, the GME concurrence is hard to compute.
In this section we derive an effective lower bound for GME concurrence in terms of the norms of the correlation tensors.

\textit{\textbf{Theorem 2:} \label{thlb}For a tripartite qudit state $\rho $, the GME concurrence satisfies the following inequality,
\begin{eqnarray}\label{T2}
C_{GME}(\rho)&\geq&\max\{\frac{1}{2\sqrt{2}}\left\| T^{(123)} \right\|-\frac{(d-1)}{d},0\}.
\end{eqnarray}}

\begin{widetext}
{\textbf{Proof.}}
We first consider pure states.
For pure state $\rho  = \left| \psi  \right\rangle \left\langle \psi  \right|$ one has $tr{\rho ^2} = 1,$ which implies that
\begin{eqnarray}\label{11}
tr{\rho ^2}= \frac{1}{{{d^3}}} + \frac{1}{{2{d^2}}} {\sum {{{(t_i^1)}^2} + } \sum {{{(t_j^2)}^2} + \sum {{{(t_k^3)}^2}} } }
+ \frac{1}{{4d}}{\sum {{{(t_{ij}^{12})}^2} + \sum {{{(t_{ik}^{13})}^2} + \sum {{{(t_{jk}^{23})}^2}} } } } + \frac{1}{8}\sum {{{(t_{ijk}^{123})}^2}}=1.
\end{eqnarray}
From (\ref{11}) we have
\[\frac{1}{{{d^3}}} + \frac{1}{{2{d^2}}}({\left\| {{T^{(1)}}} \right\|^2} + {\left\| {{T^{(2)}}} \right\|^2} + {\left\| {{T^{(3)}}} \right\|^2}) + \frac{1}{{4d}}({\left\| {{T^{(12)}}} \right\|^2} + {\left\| {{T^{(13)}}} \right\|^2} + {\left\| {{T^{(23)}}} \right\|^2}) + \frac{1}{8}{\left\| {{T^{(123)}}} \right\|^2} = 1.\]
In the following we denote $\rho_{jk}$ the reduced density matrix for the subsystems $j\neq k=1,2,3$.
One  computes from (\ref{rho}) that
$${\rho _1} = \frac{1}{d}I + \frac{1}{2}\sum {t_i^1\lambda _i^1},~~~ {\rho _{23}} = \frac{1}{d}I \otimes I + \frac{1}{{2d}}(\sum {t_j^2} \lambda _j^2 \otimes I + \sum {t_k^3} I \otimes \lambda _k^3) + \frac{1}{4}\sum {t_{jk}^{23}} {\lambda _j} \otimes {\lambda _k}.$$
Thus $$tr\rho _1^2 = \frac{1}{d} + \frac{1}{2}{\left\| {{T^{(1)}}} \right\|^2},~~~ tr\rho _{23}^2 = \frac{1}{{{d^2}}} + \frac{1}{{2d}}({\left\| {{T^{(2)}}} \right\|^2} + {\left\| {{T^{(3)}}} \right\|^2}) + \frac{1}{4}{\left\| {{T^{(23)}}} \right\|^2}.$$
Similarly we get
\begin{eqnarray*}tr\rho _2^2 &=& \frac{1}{d} + \frac{1}{2}{\left\| {{T^{(2)}}} \right\|^2},~~~tr\rho _{13}^2 = \frac{1}{{{d^2}}} + \frac{1}{{2d}}({\left\| {{T^{(1)}}} \right\|^2} + {\left\| {{T^{(3)}}} \right\|^2}) + \frac{1}{4}{\left\| {{T^{(13)}}} \right\|^2},\\
tr\rho _3^2 &=& \frac{1}{d} + \frac{1}{2}{\left\| {{T^{(3)}}} \right\|^2},~~~tr\rho _{12}^2 = \frac{1}{{{d^2}}} + \frac{1}{{2d}}({\left\| {{T^{(1)}}} \right\|^2} + {\left\| {{T^{(2)}}} \right\|^2}) + \frac{1}{4}{\left\| {{T^{(12)}}} \right\|^2}.\end{eqnarray*}
By noticing that $\rho  = \left| \psi  \right\rangle \left\langle \psi  \right|$ is a pure state, we get $tr\rho _i^2 = tr\rho _{jk}^2$ for $i \ne j \ne k$, $i,j,k = 1,2,3.$
Then we have:
\begin{eqnarray*}
\frac{3}{d}+\frac{1}{2}({\left\| {{T^{(1)}}} \right\|^2} + {\left\| {{T^{(2)}}} \right\|^2} + {\left\| {{T^{(3)}}} \right\|^2})= \frac{3}{{{d^2}}} + \frac{1}{d}({\left\| {{T^{(1)}}} \right\|^2} + {\left\| {{T^{(2)}}} \right\|^2} + {\left\| {{T^{(3)}}} \right\|^2}) + \frac{1}{4}({\left\| {{T^{(12)}}} \right\|^2} + {\left\| {{T^{(13)}}} \right\|^2} + {\left\| {{T^{(23)}}} \right\|^2}).\\
\end{eqnarray*}

Set $A = {\left\| {{T^{(1)}}} \right\|^2} + {\left\| {{T^{(2)}}} \right\|^2} + {\left\| {{T^{(3)}}} \right\|^2}$,
$B = {\left\| {{T^{(12)}}} \right\|^2} + {\left\| {{T^{(13)}}} \right\|^2} + {\left\| {{T^{(23)}}} \right\|^2}$ and
$C = {\left\| {{T^{(123)}}} \right\|^2}$. We have
\begin{equation}
\label{2}
\frac{1}{4}B = (\frac{1}{2} - \frac{1}{d})A + \frac{3}{d} - \frac{2}{{{d^2}}}.
\end{equation}
Substituting (\ref{2}) into (\ref{11}), one has
\begin{eqnarray*}
tr\rho^2&=&\frac{1}{d^3}+ \frac{1}{2d^2}A + \frac{1}{4d}B + \frac{1}{8}C=\frac{1}{d^3} + \frac{1}{2d^2}A + \frac{1}{d}\left[ {(\frac{1}{2} - \frac{1}{d})A + \frac{3}{d} - \frac{3}{{{d^2}}}} \right] + \frac{1}{8}C\\
&=&\frac{3}{d^2}-\frac{2}{d^3}+(\frac{1}{d}-\frac{1}{d^2})A + \frac{1}{8}C.
\end{eqnarray*}
Thus we get
\begin{eqnarray*}
tr\rho^2-tr\rho_1^2&=&(\frac{3}{d^2}-\frac{2}{d^3}-\frac{1}{d})+\frac{2d-2-d^2}{2d^2}||T^{(1)}||^2+(\frac{1}{d}-\frac{1}{d^2})(||T^{(2)}||^2+||T^{(3)}||^2)
+\frac{1}{8}C\\
&\geq&\frac{-d^2+3d-2}{d^3}+\frac{-d^2+2d-2}{2d^2}\frac{2(d-1)}{d}+\frac{1}{8}C=\frac{1}{8}C-\frac{(d-1)^2}{d^2},\\
\end{eqnarray*}
where we have used $||T^{(1)}||^2\leq\frac{2(d-1)}{d}$ and set $(||T^{(2)}||^2+||T^{(3)}||^2)=0$ to get the inequality.

Similarly, one obtains
\begin{eqnarray*}
tr\rho^2-tr\rho_2^2\geq\frac{1}{8}C-\frac{(d-1)^2}{d^2},~~~ tr\rho^2-tr\rho_3^2\geq\frac{1}{8}C-\frac{(d-1)^2}{d^2}.
\end{eqnarray*}
Thus we have
\begin{eqnarray*}
C_{GME}^2(|\psi\rangle)&=&\min\{1-tr\rho_1^2,1-tr\rho_2^2,1-tr\rho_3^2\}=\min\{tr\rho^2-tr\rho_1^2,tr\rho^2-tr\rho_2^2,tr\rho^2-tr\rho_3^2\}\\
&\geq&\max\{\frac{1}{8}C-\frac{(d-1)^2}{d^2},0\}.
\end{eqnarray*}

We now consider mixed quantum state $\rho\in H_1^d \otimes H_2^d \otimes H_3^d.$
Let $\rho= \sum {{p_\alpha }} \left| {{\psi _\alpha }} \right\rangle \left\langle {{\psi _\alpha }} \right| $ be the optimal ensemble decomposition of $\rho$. We obtain
\begin{eqnarray*}
C_{GME}(\rho)=\sum p_{\alpha}C_{GME}(|\psi_{\alpha}\rangle)\geq\max\{(\frac{1}{2\sqrt{2}}\left\| T^{(123)}_{\alpha} \right\|-\frac{(d-1)}{d}),0\}\sum p_{\alpha}=\max\{\frac{1}{2\sqrt{2}}\left\| T^{(123)} \right\|-\frac{(d-1)}{d},0\},\\
\end{eqnarray*}
where we have used $\sqrt{a-b}\geq\sqrt{a}-\sqrt{b}$ for $a>b>0$ and the convex property of the Frobenius norm.
\hfill \rule{1ex}{1ex}\end{widetext}

\textit{\textbf{Example 2: }}\label{ex31} We consider the mixture of the GHZ state and W state in three-qubit quantum systems $\rho= \frac{1-x-y}{8}I + x|GHZ\rangle\langle GHZ|+y|W\rangle\langle W|,$ where $|GHZ\rangle= \frac{1}{\sqrt 2}(|000\rangle  + |111\rangle)$ and $|W\rangle= \frac{1}{\sqrt 3}(|001\rangle  + |010\rangle+|100\rangle)$. The lower bound of GME concurrence for $\rho$ by Theorem 2 is computed to be $\max\{\frac{1}{12}(\sqrt{72x^2+66y^2}-6),0\}$ as shown in Figure 2. Comparing with Vicente's criterion, our lower bound can detect more genuine multipartite entangled states as shown in Figure 3.
\begin{figure}[h]
\begin{center}
\resizebox{6cm}{!}{\includegraphics{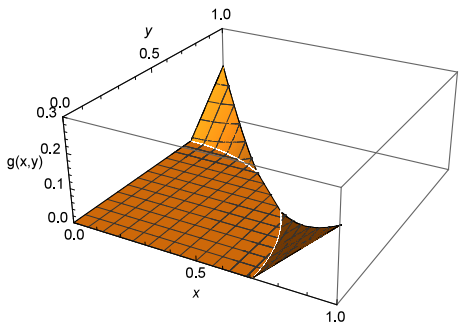}}
\end{center}
\caption{Lower bound of GME concurrence in Theorem 2 for $\rho$ in Example 2. $g(x,y)$ stands for the lower bound.\label{fig3}}
\end{figure}
\begin{figure}[h]
\begin{center}
\resizebox{6cm}{!}{\includegraphics{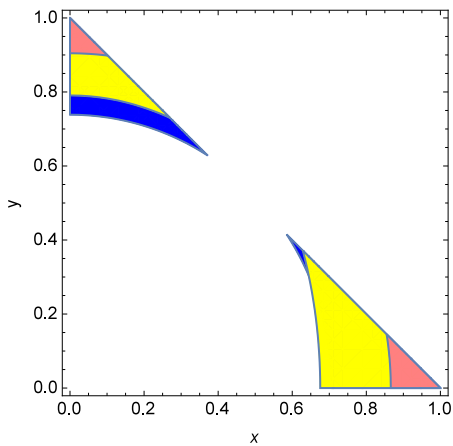}}
\end{center}
\caption{GME Detected by vicente criterion (pink region by Theorem 1 and yellow region by Theorem 2 in \cite{vicente3}) and by the lower bound for GME concurrence in our Theorem 2 (blue region).\label{fig3}}
\end{figure}

\textit{\textbf{Example 3: }} We consider a mixed state in three-qutrit quantum systems $\rho= \frac{1-x}{27}I + x|\psi\rangle\langle \psi|,$ where $|\psi\rangle= \frac{1}{\sqrt 3}(|012\rangle  + |021\rangle+ |111\rangle)$. The lower bound of GME concurrence can detect GME better than Vicente's criterion(Theorem 1) and Theorem 1 in this manuscript as shown in Figure 4.

\begin{figure}[h]
\begin{center}
\resizebox{6cm}{!}{\includegraphics{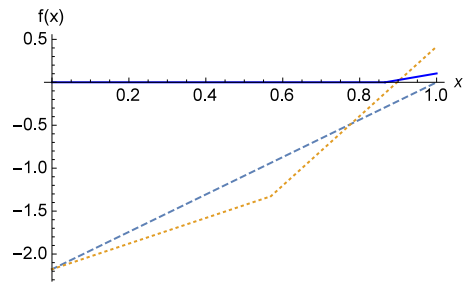}}
\end{center}
\caption{GME detection: dashed line by Theorem 1 in \cite{vicente3}), dotted and solid line by our Theorem 1 and Theorem 2 respectively. $f(x)$ denotes the difference between the left and the right side of the inequalities in these criteria.
We compute that $f(x)=2.17732 (-1 + x), -3.628874 + 4.044882x$ and $\max\{0.222222 (-3 + 3.4641x),0\}$ for Theorem 1 in \cite{vicente3},  our Theorem 1 and Theorem 2 respectively. One finds from the figure that Theorem 1 in \cite{vicente3} can not detect GME for the whole region of $x$ in this example, Theorem 2 in \cite{vicente3} can not be operated on this states since it only fits for tripartite qubits system, while Theorem 1 and Theorem 2 in this manuscript detect GME for $0.9<x\leq 1$ and $0.866025<x\leq 1$ respectively. \label{fig3}}
\end{figure}

\section{Conclusions and Discussions}\label{sec4}
It is a basic and fundamental question in quantum theory to detect and quantify GME.
Since the GME concurrence is defined by optimization
over all ensemble decompositions of a mixed quantum state, it
is a formidable task to derive an analytical formula. We have derived an analytical and
experimentally feasible lower bound for GME concurrence
of any tripartite quantum state based on the correlation tensors of the density matrix.
We have also obtained an effective criterion to detecting GME for any tripartite quantum states by sketching the Vicente's method.
Genuine multipartite entanglement can be detected by
using this bound. The results presented in this manuscript are experimentally feasible as the elements in the correlation tensors are
just the mean values of the Hermitian $SU(d)$ generators. The approach used in this manuscript
can also be implemented to investigate the k-separability of multipartite
quantum systems. The results can be also generalized to any multipartite qudit systems.

\bigskip
\noindent{\bf Acknowledgments}\, \, This work is supported by the NSFC No.11775306, No.11701568, No.11675113; the Fundamental Research Funds for the Central Universities Grants No. 15CX05062A, No. 16CX02049A, and 17CX02033A; the Shandong Provincial Natural Science Foundation No.ZR2016AQ06 and No. ZR2017BA019; Qingdao applied basic research program No. 15-9-1-103-jch, the NSF of China under Grant No. 11675113, and a project sponsored by SRF for ROCS, SEM.


\smallskip
\begin{thebibliography}{99}
\bibitem{nielsen}M.A. Nielsen and I.L. Chuang, Quantum Computation and Quantum
Information. Cambridge: Cambridge University Press, (2000).

\bibitem{di} See, for example, D.P. Di Vincenzo, Science
270,255 (1995).

\bibitem{guhnerev} O. Guhne, and G. Toth, Physics Reports 474, 1 (2009).

\bibitem{mule1} R. Horodecki et al., Rev. Mod. Phys. 81, 865 (2009).

\bibitem{mule2} H.J. Briegel, D.E. Browne, W. D$\ddot{u}$r, R. Raussendorf and M. Van den Nest, Nat. Phys. 5, 19 (2009).

\bibitem{mule3} A. Sen(De) and U. Sen, Phys. News 40, 17 (2010).
arXiv:1105.2412.

\bibitem{mule4} N. Gisin, G. Ribordy, W. Tittel, and H. Zbinden, Rev. Mod. Phy. 74, 145 (2002).

\bibitem{mule5} R. Cleve et al., Phys. Rev. Lett. 83, 648 (1999); A. Karlsson et
al., Phys. Rev. A 59, 162 (1999).

\bibitem{srensen} A. S. Srensen and K. Mlmer, Phys. Rev. Lett. 86, 4431
(2001).
\bibitem{hyllus} P. Hyllus,W. Laskowski, R. Krischek, C. Schwemmer, W.
Wieczorek, H. Weinfurter, L. PezzAe, and A. Smerzi,
Phys. Rev. A 85, 022321 (2012).
\bibitem{rauss} R. Raussendorf and H.J. Briegel, Phys. Rev. Lett 86,
5188 (2001).

\bibitem{murao} M. Murao, D. Jonathan, M.B. Plenio and V. Vedral,
Phys. Rev. A 59, 156-161 (1999).
\bibitem{hillery} M. Hillery, V. Buzek, A. Berthiaume, Phys. Rev. A. 59,
1829 (1999).
\bibitem{scarani} V. Scarani and N. Gisin, Phys. Rev. Lett. 87, 117901
(2001).
\bibitem{zhao} Z. Zhao, Y.A. Chen, A.N. Zhang, T. Yang, H.J. Briegel
and J.W. Pan, Nature 430, 54 (2004).

\bibitem{telem} Y. Yeo and W.K. Chua, Phys. Rev. Lett. 96, 060502
(2006), P.X. Chen, S.Y. Zhu, and G.C. Guo, Phys. Rev. A 74, 032324
(2006).

\bibitem{12} M. Huber, F. Mintert, A. Gabriel, and
B. C. Hiesmayr, Phys. Rev. Lett. 104, 210501 (2010).

\bibitem{huber} M. Huber and R. Sengupta, Phys. Rev. Lett. 113, 100501
(2014).

\bibitem{vicente3} J.I. de Vicente, M. Huber, Phys. Rev. A, 84, 062306
(2011).

\bibitem{wu} J.Y. Wu, H. Kampermann, D. Bru${\ss}$, C. Klockl, and M. Huber,
Phys. Rev. A 86, 022319 (2012).

\bibitem{huber1} M. Huber, M. Perarnau-Llobet, J.I.
de Vicente, Phys. Rev. A 88, 042328 (2013).

\bibitem{sperling} J. Sperling, W. Vogel, Phys. Rev. Lett. 111, 110503 (2013).



\bibitem{14} C. Eltschka and J. Siewert, Phys. Rev. Lett. 108, 020502
(2012).

\bibitem{15} B. Jungnitsch, T. Moroder, and O. G$\ddot{u}$hne, Phys. Rev. Lett. 106,
190502 (2011).

\bibitem{monogamy} V. Coffman, J. Kundu, and W. K. Wootters, Phys. Rev. A
61, 052306 (2000); T.J. Osborne and F. Verstraete, Phys. Rev. Lett.
96, 220503 (2006); B. Regula, S.D. Martino, S. Lee, and G. Adesso,
Phys. Rev. Lett. 113, 110501 (2014).


\bibitem{claude} C. Klckl, M. Huber, Phys. Rev. A 91, 042339
(2015).

\bibitem{horo} M. Markiewicz, W. Laskowski, T. Paterek, and M. $\dot{Z}$ukowski
Phys. Rev. A 87, 034301 (2013).

\bibitem{ma1} Z.H. Ma, Z.H. Chen, J.L. Chen, C.
Spengler, A. Gabriel, and M. Huber, Phys. Rev. A 83, 062325 (2011).

\bibitem{ma2} Z.H. Chen, Z.H. Ma, J.L. Chen, and S. Severini, Phys. Rev. A 85, 062320
(2012).

\bibitem{gaot1} Y. Hong, T. Gao, and F.L. Yan, Phys. Rev. A 86, 062323 (2012).

\bibitem{gaot2} T. Gao,. F.L. Yan, and S.J. van Enk, Phys. Rev. Lett. 112, 180501
(2014).

\bibitem{bellgme} J.D. Bancal, N. Gisin, Y.C. Liang, and S. Pironio, Phys. Rev.
Lett. 106, 250404 (2011).

\bibitem{10} C. Lancien, O. Guhne, R. Sengupta, M. Huber, J. Phys. A: Math. Theor. 48 505302 (2015).

\bibitem{wootters} W.K. Wootters, Phys. Rev. Lett. 80, 2245 (1998).

\bibitem{multicon1} L. Aolita and F. Mintert, Phys. Rev. Lett. 97,
050501 (2006).

\bibitem{multicon2} A.R.R. Carvalho, F. Mintert, and A. Buchleitner, Phys. Rev. Lett.
93, 230501 (2004).

\bibitem{chenk} K. Chen, S. Albeverio and S.M. Fei, Phys. Rev. Lett. 95,
040504 (2005).

\bibitem{vicente1} J.I. de Vicente, Quantum Inf. Comput. 7, 624(2007).


\bibitem{vicente2} J.I. de Vicente, J. Phys. A: Math. and Theor., 41, 065309 (2008).


\bibitem{hassan} A.S. M. Hassan, P. S. Joag, Quant. Inf. Comput. 8, 0773 (2008).

\bibitem{ming} M. Li, J. Wang, S.M. Fei and X.Q. Li-Jost, Phys. Rev. A, 89,022325 (2014).

\bibitem{hassan1} A.S. M. Hassan, P. S. Joag, Phys. Rev. A, 77, 062334
(2008).


\bibitem{hassan2} A.S. M. Hassan, P. S. Joag, Phys. Rev. A, 80, 042302
(2009).


\bibitem{mingbell} M. Li and S.M. Fei, Phys. Rev. A, 86, 052119
(2012).

\bibitem{GE}M. Li, J. Wang, S.M. Fei, X.Q. Li-Jost, and H. Fan, Phys. Rev. A 92, 062338 (2015).

\bibitem{lambda} G. Kimura, Phys. Lett. A, 314, 339(2003).

\end{thebibliography}
\end{document}